\documentclass[aps,prl,twocolumn,amsmath,superscriptaddress]{revtex4}
\pdfoutput=1
\usepackage{graphicx}
\usepackage{eufrak}
\newcommand{\be}{\begin{equation}}
\newcommand{\ee}{\end{equation}}

\begin{document}

\title{Realization of an optomechanical interface between ultracold atoms and a membrane}

\author{Stephan Camerer}\thanks{These authors contributed equally to this work.}
\affiliation{Fakult{\"a}t f{\"u}r Physik, Ludwig-Maximilians-Universit{\"a}t, Schellingstra{\ss}e 4, 80799 M{\"u}nchen, Germany}
\affiliation{Max-Planck-Institut f{\"u}r Quantenoptik,  Hans-Kopfermann-Str. 1, 85748 Garching, Germany}

\author{Maria Korppi}\thanks{These authors contributed equally to this work.}
\affiliation{Fakult{\"a}t f{\"u}r Physik, Ludwig-Maximilians-Universit{\"a}t, Schellingstra{\ss}e 4, 80799 M{\"u}nchen, Germany}
\affiliation{Max-Planck-Institut f{\"u}r Quantenoptik,  Hans-Kopfermann-Str. 1, 85748 Garching, Germany}
\affiliation{Departement Physik, Universit{\"a}t Basel, Klingelbergstrasse 82, 4056 Basel, Switzerland}

\author{Andreas J\"{o}ckel}
\affiliation{Fakult{\"a}t f{\"u}r Physik, Ludwig-Maximilians-Universit{\"a}t, Schellingstra{\ss}e 4, 80799 M{\"u}nchen, Germany}
\affiliation{Max-Planck-Institut f{\"u}r Quantenoptik,  Hans-Kopfermann-Str. 1, 85748 Garching, Germany}
\affiliation{Departement Physik, Universit{\"a}t Basel, Klingelbergstrasse 82, 4056 Basel, Switzerland}

\author{David Hunger}
\affiliation{Fakult{\"a}t f{\"u}r Physik, Ludwig-Maximilians-Universit{\"a}t, Schellingstra{\ss}e 4, 80799 M{\"u}nchen, Germany}
\affiliation{Max-Planck-Institut f{\"u}r Quantenoptik,  Hans-Kopfermann-Str. 1, 85748 Garching, Germany}
\author{Theodor W. H\"{a}nsch}
\affiliation{Fakult{\"a}t f{\"u}r Physik, Ludwig-Maximilians-Universit{\"a}t, Schellingstra{\ss}e 4, 80799 M{\"u}nchen, Germany}
\affiliation{Max-Planck-Institut f{\"u}r Quantenoptik,  Hans-Kopfermann-Str. 1, 85748 Garching, Germany}

\author{Philipp Treutlein}\email[To whom correspondence should be addressed. Electronic address: ]{philipp.treutlein@unibas.ch}
\affiliation{Fakult{\"a}t f{\"u}r Physik, Ludwig-Maximilians-Universit{\"a}t, Schellingstra{\ss}e 4, 80799 M{\"u}nchen, Germany}
\affiliation{Max-Planck-Institut f{\"u}r Quantenoptik,  Hans-Kopfermann-Str. 1, 85748 Garching, Germany}
\affiliation{Departement Physik, Universit{\"a}t Basel, Klingelbergstrasse 82, 4056 Basel, Switzerland}

\date{\today}

\begin{abstract}
We have realized a hybrid optomechanical system by coupling ultracold atoms to a micromechanical membrane. 
The atoms are trapped in an optical lattice, which is formed by retro-reflection of a laser beam from the membrane surface. In this setup, the lattice laser light mediates an optomechanical coupling between membrane vibrations and atomic center-of-mass motion.
We observe both the effect of the membrane vibrations onto the atoms as well as the backaction of the atomic motion onto the membrane. 
By coupling the membrane to laser-cooled atoms, we engineer the dissipation rate of the membrane. 
Our observations agree quantitatively with a simple model.
\end{abstract}


\maketitle

Laser light can excert a force on material objects through radiation pressure and through the optical dipole force \cite{Chu1991}. 
In the 
very active field of optomechanics \cite{optomech_reviews}, such light forces are exploited for cooling and control of the vibrations of mechanical oscillators, with possible applications in precision force sensing and studies of quantum physics at macroscopic scales.
This has many similarities with the field of ultracold atoms \cite{chu_insight}, where radiation pressure forces are routinely used for laser cooling \cite{Chu1991} and optical dipole forces are used for trapping and quantum manipulation of atomic motion, most notably in optical lattices \cite{Bloch2005,Murch2008}.

In a number of recent theoretical papers it has been proposed that light forces could also be used to couple the motion of atoms in a trap to the vibrations of a single mode of a mechanical oscillator \cite{prop4, prop6, prop7, Chang09, prop9, prop10, strong_single_atom, Wallquist2010, optlattmirror, prop11, Hunger2011}. 
In the resulting hybrid optomechanical system the atoms could be used to read out the motion of the oscillator, to engineer its dissipation, and ultimately to perform quantum information tasks such as coherently exchanging the quantum state of the two systems. 
In recent experiments using magnetic \cite{kitching} or surface-force coupling \cite{hunger_prl}, atoms were used to detect vibrations of micromechanical oscillators. However, the backaction of the atoms onto the oscillator vibrations, which is required for cooling and manipulating the oscillator with the atoms, could not be observed.

Here we report the experimental implementation of a hybrid optomechanical system in which an optical lattice mediates a long-distance coupling between ultracold atoms and a micromechanical membrane oscillator \cite{optlattmirror}. 
If the trap frequency of the atoms in the lattice is matched to the eigenfrequency of the membrane, the coupling leads to resonant energy transfer between the two systems. 
We observe both the effect of the membrane vibrations onto the atoms as well as the backaction of the atomic motion onto the membrane. 
We demonstrate that the dissipation rate of the membrane can be engineered by coupling it to laser-cooled atoms, as predicted by recent theoretical work \cite{optlattmirror}. 

\begin{figure}
\centering
\includegraphics[width=1\columnwidth]{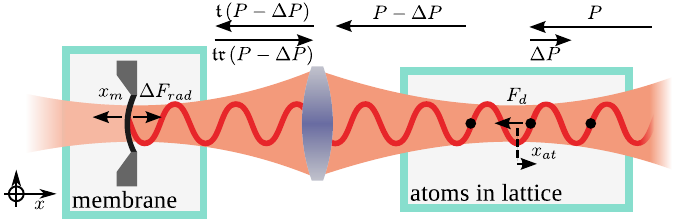}
\caption{\label{fig:paper1} Optomechanical coupling of atoms and membrane. A laser beam of power $P$ 
is partially reflected at a SiN membrane of reflectivity $\mathfrak{r}$ and forms a 1D optical lattice for an ultracold atomic ensemble. Motion of the membrane displaces the lattice and thus couples to atomic motion. Conversely, atomic motion is imprinted as a power modulation $\Delta P$ onto the laser, thus modulating the radiation pressure force on the membrane. 
$\mathfrak{t}$ is the transmittivity of the optics between atoms and membrane. 
Arrows illustrate the direction of forces and displacements at a specific point in time. In the main text, all forces and displacements are positive if pointing to the right.
}
\end{figure}


The coupling scheme we investigate 
is illustrated in Fig.~\ref{fig:paper1}, see also \cite{optlattmirror}. 
A laser beam of power $P$, whose frequency $\omega$ is red detuned with respect to an atomic transition, impinges from the right onto a SiN membrane  oscillator and is partially retroreflected. The reflected beam is overlapped with the incoming beam such that a 1D optical lattice potential for ultracold atoms is generated \cite{Bloch2005}. 
A displacement of the membrane $x_m$ displaces the lattice potential, resulting in a force $F=m\omega_{at}^2x_m$ onto each atom, where $m$ is the atomic mass and $\omega_{at}$ the trap frequency in a harmonic approximation to the lattice potential well. The membrane motion thus couples through $F_{com} = NF$ to the center of mass (c.o.m.)\ motion of an ensemble of $N$ atoms trapped in the lattice. 
Conversely, an atom displaced by $x_{at}$ from the bottom of its potential well experiences a restoring optical dipole force $F_d=-m \omega_{at}^2 x_{at}$ in the lattice. 
On a microscopic level, $F_d$ 
is due to absorption and stimulated emission, leading to a redistribution of photons between the two running wave components forming the lattice \cite{raithelgoerlitz}. Each redistribution event results in 
a momentum transfer of $\pm 2 \hbar k$ to the atom, where $k=\omega/c$. The photon redistribution modulates the power of the laser beam traveling towards the membrane by $\Delta P=\hbar \omega \dot{n}=-(c/2)NF_d$, where $\dot{n}$ is the total redistribution rate due to the $N$ atoms. 
The atomic c.o.m.\ motion is thus imprinted onto the laser light. The resulting modulation $\Delta F_{rad}=(2/c)\mathfrak{r} \mathfrak{t}  \Delta P$ of the radiation pressure force exerted by the laser on the membrane constitutes the backaction of the atoms.

In a simple model of harmonic oscillators coupled through $F_{com}$ and $\Delta F_{rad}$, the equations of motion for the fundamental vibrational mode of the membrane and the c.o.m.\ motion of the atoms can be written as
\begin{eqnarray*}
\dot{p}_{at} &=& -\gamma_{at} p_{at} - N m \omega_{at}^2 x_{at} + N m \omega_{at}^2 x_m  \\
\dot{x}_{at}&=& p_{at} / N m\\
\dot{p}_{m} &=& -\gamma_m p_m - M \omega_{m}^2 x_{m} + \mathfrak{r}\mathfrak{t} N m \omega_{at}^2 x_{at}  \\
\dot{x}_{m}&=& p_{m} / M
\end{eqnarray*}
where $\omega_m$ and $M$ are frequency and effective mass of the membrane mode, 
and $\gamma_m$ ($\gamma_{at}$) is the motional damping rate of the membrane (atoms). 
We introduce dimensionless complex amplitudes 
$a = e^{i \omega_m t} \sqrt{Nm\omega_{at} / 2\hbar}\left( x_{at} + i p_{at} / Nm\omega_{at}\right)$ and
$b = e^{i \omega_m t} \sqrt{M\omega_{m} / 2\hbar}\left( x_{m} + i p_{m} / M\omega_{m}\right)$ 
in a frame rotating at  $\omega_m$ 
and rewrite the equations of motion in the rotating-wave approximation as
\begin{eqnarray}
\dot{a} &=& - i \delta a - (\gamma_{at}/2) a + i g b,  \nonumber \\
\dot{b} &=& - (\gamma_{m}/2) b + i \mathfrak{r} \mathfrak{t} g a. \label{eq:ab}
\end{eqnarray}
%
%
Here, $g=\frac{\omega_{at}}{2}\sqrt{\frac{N m \omega_{at}}{M\omega_m}}$ is the coupling constant and $\delta=\omega_{at}-\omega_m$ the detuning. 
We note that the coupling between atoms and membrane is asymmetric. 
Some of the photons that have interacted with the atoms are lost because $\mathfrak{r}\mathfrak{t}<1$ and do not contribute to the force on the membrane. For $\mathfrak{r}\mathfrak{t}=1$, symmetric coupling is recovered, as expected from the actio-reactio principle. 
A full quantum theory of our system confirms these results \cite{optlattmirror}. 

The atomic damping rate $\gamma_{at} = \gamma_c + \gamma_\phi$ can be manipulated by applying laser cooling at rate $\gamma_c$ to the atoms. 
It also accounts for additional dephasing of the c.o.m.\ motion at rate $\gamma_\phi$. 
In our experimental realization $\gamma_{at}\gg g, \gamma_m$ so that the atomic c.o.m.\ amplitude is approximately in steady state ($\dot{a}\simeq 0$) on the much slower timescale of membrane dynamics.
In this regime, initially excited membrane vibrations decay according to Eqs.~(\ref{eq:ab}) as $|b(t)|^2 = |b(0)|^2 \exp(-\Gamma t)$, with a decay rate
\begin{equation}\label{eq:Gamma}
\Gamma = \gamma_m+ \gamma_{at} \frac{ g^2 \mathfrak{r} \mathfrak{t}}{\delta^2 + (\gamma_{at}/2)^2}.
\end{equation}
The second term in Eq.~(\ref{eq:Gamma}) is an additional dissipation rate that arises due to the membrane's coupling to laser-cooled atoms in the lattice.
In the following, we describe experiments where we observe and study this effect. 

\begin{figure}
\centering
\includegraphics[width=0.95\columnwidth]{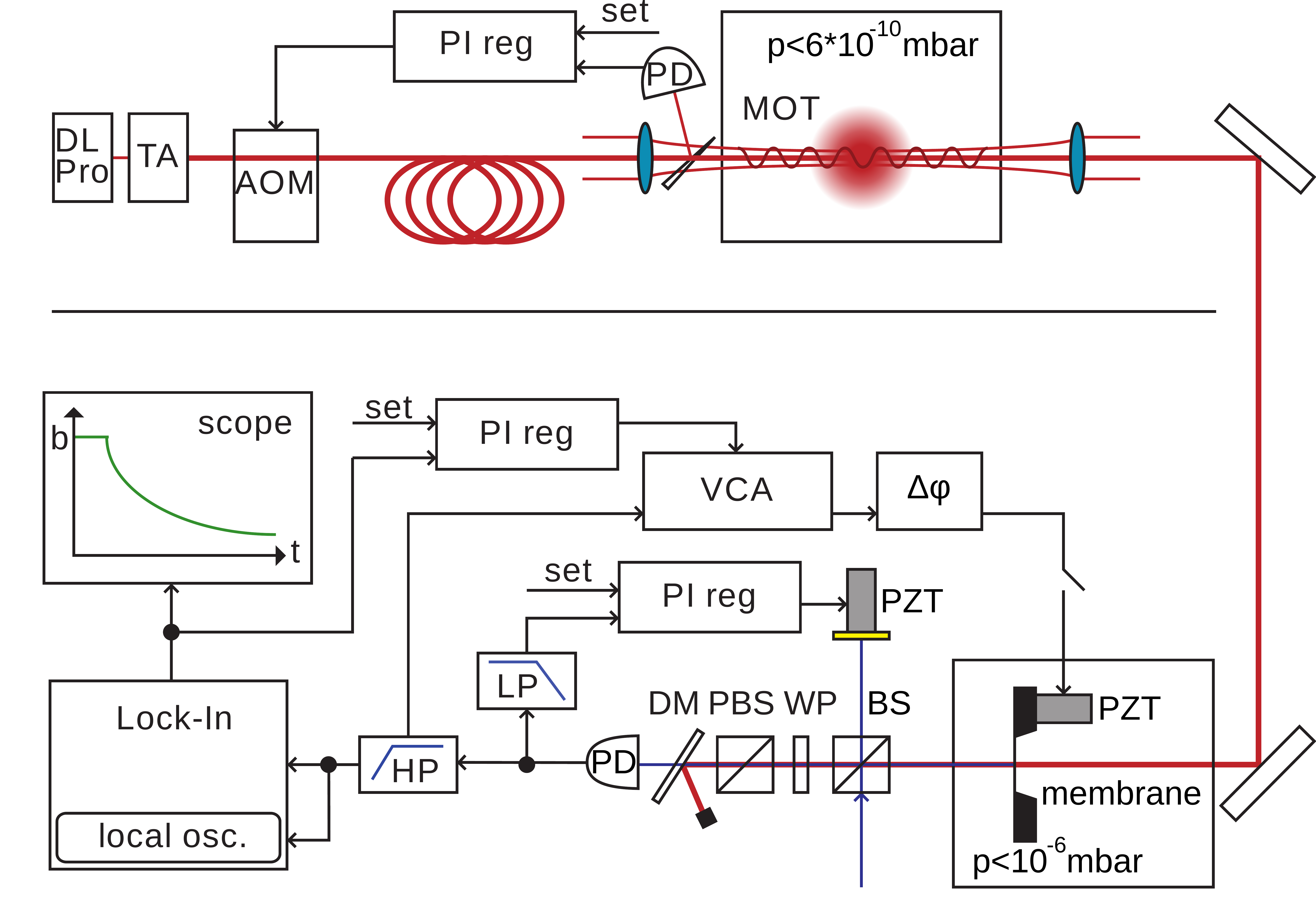}
\caption{\label{fig:paper2} Experimental setup. The lattice laser (red) is fiber coupled, power stabilized with a PI regulator and an acousto-optic modulator (AOM), and focused into the MOT vacuum chamber. The membrane in a second room-temperature vacuum chamber serves as partially reflective end mirror for the 1D optical lattice. The membrane motion is read out with a Michelson interferometer (blue laser). The two lasers are separated with a $\lambda/2$-plate (WP), a polarizing beam spitter (PBS) and a dichroic mirror (DM). The interferometer signal from the photodetector (PD) is frequency-split: the low-frequency part (LP) is used for interferometer stabilization; the high-frequency part (HP) including the membrane signal is used for readout and a piezo (PZT) feedback drive of the membrane. The membrane amplitude is measured with a lock-in amplifier and an oscilloscope. When driven, it is stabilized with a PI regulator and a voltage controlled amplifier (VCA) in the feedback loop ($\Delta\varphi$: phase shift).
}
\end{figure}

Our experimental setup is sketched in Fig.~\ref{fig:paper2}. The 1D optical lattice potential is provided by a grating-stabilized diode laser (DL pro) injecting a tapered amplifier (TA). The laser frequency is red detuned by $\Delta=-2 \pi \times 21$~GHz from the $D_2$ line of $^{87}$Rb ($F=2 \leftrightarrow F'=3$ transition). The power $P$ is actively stabilized to a relative stability of $2 \times 10^{-4}$ r.m.s.\ in a bandwidth of $12$~kHz. 
At the position of the atoms, $P$ can be adjusted in the range of $0...140$~mW. 
The linearly polarized lattice beam is sent through a vacuum chamber with ultracold $^{87}$Rb atoms and is partially reflected at the surface of a SiN membrane mounted in a separate vacuum chamber. 
The reflected and incoming laser beams form an optical lattice with a beam waist $w_0= 350 \pm 30~\mu$m at the position of the atoms. The lattice potential is only partially modulated because the reflected beam is weaker than the incoming beam ($\mathfrak{r}=0.28$, $\mathfrak{t}=0.82$).  
For an incoming beam of $P=76$~mW, the calculated modulation depth of the sinusoidal potential is $V_0 = k_B \times 290\pm 50~\mu$K and $\omega_{at}/2\pi = 305\pm25$~kHz \cite{Grimm2000}. 
By changing $P$, we change $\omega_{at} \propto \sqrt{P}$ and $V_0 \propto P$.  
The lattice is loaded with laser-cooled atoms from a magneto-optical trap (MOT) \cite{mirrormot}. 
At $P=76$~mW, we typically load $N=2\times 10^6$ atoms into the lattice with a temperature of $T=100~\mu$K. 

The SiN membrane \cite{Zwickl08} has a tensile stress of about $120$~MPa, dimensions of $0.5~\mathrm{mm}\times 0.5~\mathrm{mm}\times50$~nm, $\mathfrak{r}=0.28$ at $\lambda=780$~nm, and a fundamental vibrational mode with $\omega_m / 2 \pi=272$~kHz and $M=1 \times 10^{-11}$~kg. 
The membrane vibrations are read out with a Michelson interferometer with a 
position sensitivity of $3 \times 10^{-14}$~m$/\sqrt{\textrm{Hz}}$. We observe that $\omega_m$ decreases with increasing lattice laser power, and measure $\omega_m / 2 \pi=244$~kHz at $P=76$~mW. 
We attribute this to reduced tensile stress due to thermal expansion of the membrane, which is locally heated by the lattice laser \cite{tobepub}.
The mechanical quality factor $Q=\omega_m/\gamma_m= \omega_m \tau / 2$ of the fundamental mode is determined in ringdown measurements from the $1/e$ decay time $\tau$ of the initially excited membrane amplitude. 
We find $Q=8.5 \times 10^5$ ($Q=1.5\times 10^6$) for $P=0$ ($P=76$~mW). We observe that $Q$ changes reproducibly in a non-linear way with $P$ \cite{tobepub}. 
In the following experiments, we measure at values of $P$ where the r.m.s.\ fluctuations of $\gamma_m$ are below $0.015~\mathrm{s}^{-1}$. 

\begin{figure}
\centering
\includegraphics[width=0.9\columnwidth]{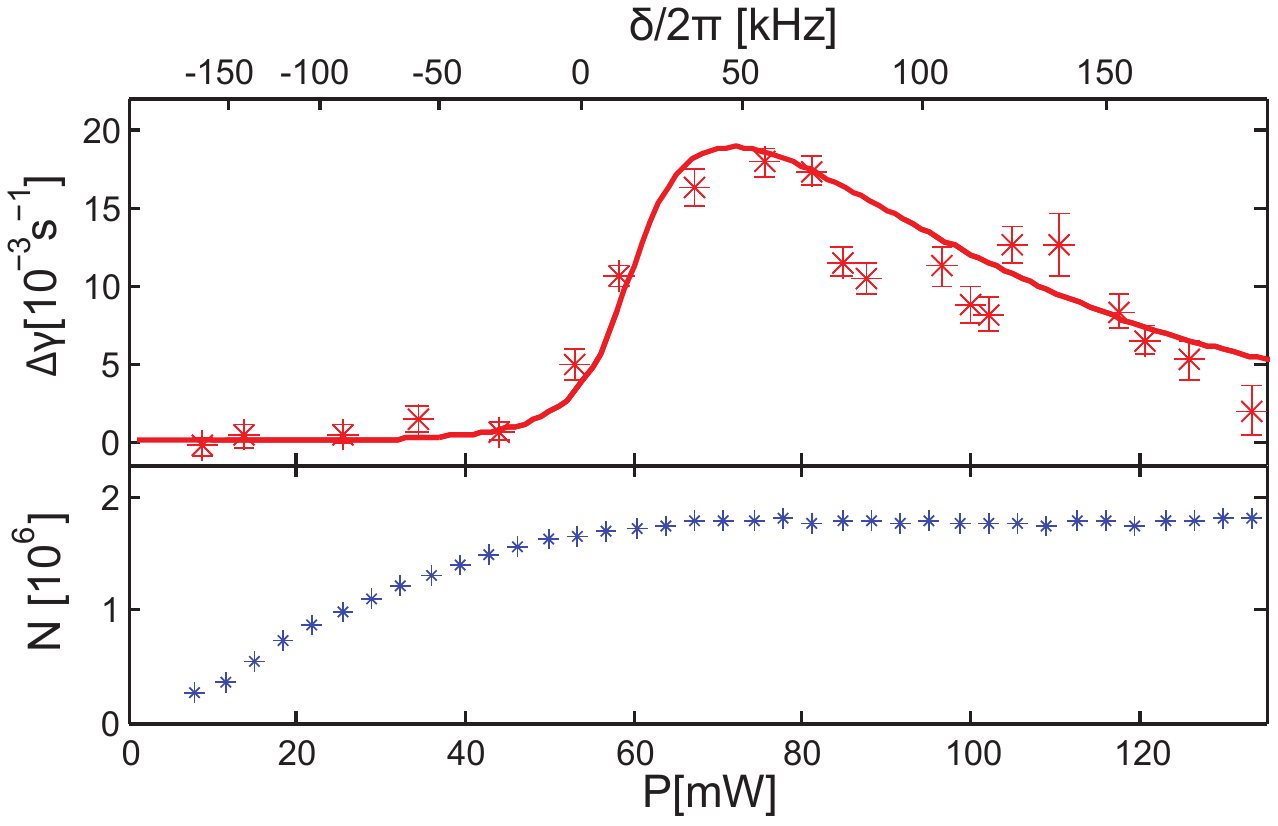}
\caption{\label{fig:paper4}Backaction of laser-cooled atoms onto the membrane. Top: measured additional membrane dissipation rate $\Delta \gamma=\Gamma-\gamma_{m}$ due to coupling to atoms as a function of $P$. The rates $\Gamma$ and $\gamma_m$ are extracted from exponential fits to averaged decay curves ($2 \times 455$ experimental runs per datapoint). Solid line: theory for a thermal ensemble in the lattice (see text). Bottom: lattice atom number in the experiment.}
\end{figure}

The backaction of the laser-cooled atomic ensemble onto the membrane vibrations is observed in membrane ringdown measurements. 
While the lattice is continuously loaded from the MOT, the membrane is resonantly excited to an amplitude of $540$~pm. 
After switching the excitation off, the decay of the membrane vibrations is recorded. 
The strong MOT laser cooling ensures that the atomic ensemble is in steady-state throughout the experiment. 
We perform alternating experiments with and without atoms in the lattice and determine the respective membrane decay rates $\Gamma$ and $\gamma_m$. The presence of atoms is controlled by detuning the MOT laser frequency, otherwise the sequences are identical. 
The measured additional membrane dissipation rate $\Delta \gamma=\Gamma-\gamma_{m}$ due to the atoms is shown in Fig.~\ref{fig:paper4} as a function of $P$. 
By varying $P$, we vary $\delta (P) = \omega_{at} (P) - \omega_{m} (P)$, where $\omega_{at} (P)$ is calculated and $\omega_{m} (P)$ is measured (see above). 
We observe a broad resonance in $\Delta\gamma$ at $P\approx 76$~mW. 
The resonance is broadened and shifted to $\delta >0$ because of finite $T$ of the atoms, leading to a variation of $\omega_{at}$ across the ensemble in the transverse intensity profile of the lattice laser (see below).  
The atom number $N$ does not vary significantly around the resonance. 
Despite the enormous mass difference $Nm/M \simeq 10^{-8}$, we clearly observe the effect of the atoms on the membrane.

\begin{figure}
\centering
\includegraphics[width=0.9\columnwidth]{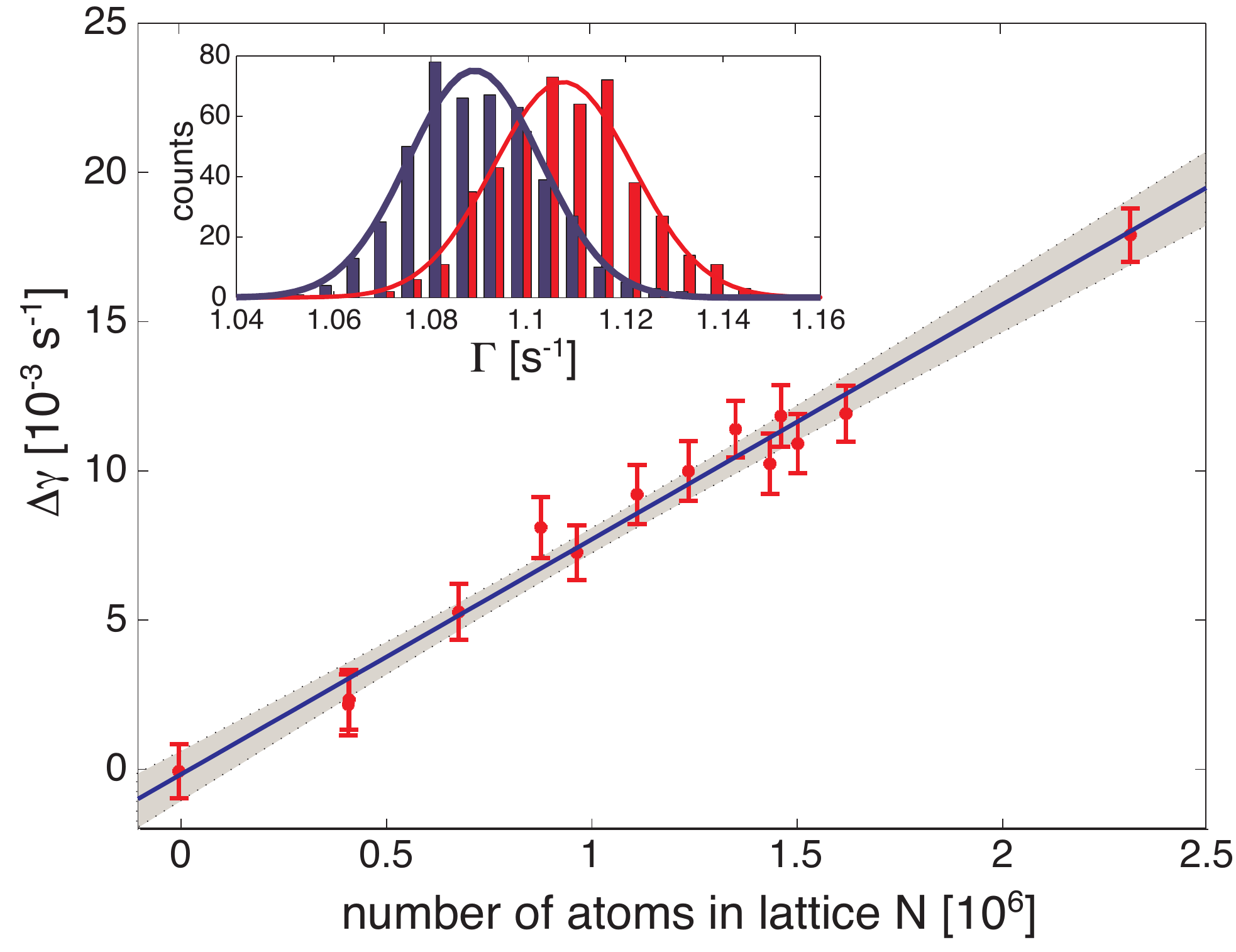}
\caption{\label{fig:paper5}Measured additional membrane dissipation $\Delta \gamma$ as a function of atom number for resonant coupling ($P=76$~mW). The blue line is a linear fit. The observed dependence agrees well with theory. Inset: histogram of measurements of $\Gamma$ for $N=2.3\times10^6$ (red) and $N=0$ (blue).}
\end{figure}

In order to study the dependence of the membrane dissipation on atom number, the system is prepared on resonance ($P=76$~mW) and $N$ is varied by varying the power of the MOT repump laser. We observe a linear dependence of $\Delta \gamma$ on $N$, see Fig.~\ref{fig:paper5}. This agrees with Eq.~(\ref{eq:Gamma}) as well as the theory in \cite{optlattmirror}. 
In order to compare measurement and theory, we calculate $\Delta \gamma$ from Eq.~(\ref{eq:Gamma}) with the overall atomic damping rate taken as the FWHM of the resonance in Fig.~\ref{fig:paper4}, $\gamma_{at}=2\pi \times 130$~kHz. For $N=2.3 \times 10^6 $ the theory predicts $\Delta \gamma=0.023\pm 0.005 $~s$^{-1}$, assuming errors of $20\%$ on $N$ and $\gamma_{at}$. This is to be compared to the measured value of $\Delta \gamma=0.018\pm 0.001$~s$^{-1}$ in Fig.~\ref{fig:paper5}. The quantitative agreement of measurement and theory is rather remarkable, as the simple model presented above does not explicitly account for finite $T$, lattice trap anharmonicity, and the spatial variation of $\omega_{at}$. 
These effects are only implicitly included in the measured $\gamma_{at}$. For our MOT detuning of $28$~MHz (including the light shift of the lattice) we estimate a laser cooling rate of $\gamma_{c}=2\pi\times 30$~kHz, implying $\gamma_\phi = 2\pi\times 100$~kHz.

In a more sophisticated model, we describe the atoms by a thermal density distribution $n(\mathbf{r})$ of constant $T=100~\mu$K in the lattice potential.
For each atom in the distribution, we calculate $\omega_{at}(\mathbf{r})$ from $V_0(\mathbf{r})$ and determine the corresponding membrane damping rate as in Eq.~(\ref{eq:Gamma}), but with $N=1$. We set $\gamma_{at}=\gamma_c$, as the effects contributing to $\gamma_\phi$ are now explicitly modeled.  We then add up the damping rates of all the atoms in the ensemble. 
The resulting line in Fig.~\ref{fig:paper4} shows good agreement with the data for $N=2.0\times 10^6$ and $w_0 = 370~\mu$m, within the uncertainty of these parameters. This shows that finite $T$ is responsible for the observed shape of the resonance.

\begin{figure}
\centering
\includegraphics[width=0.9\columnwidth]{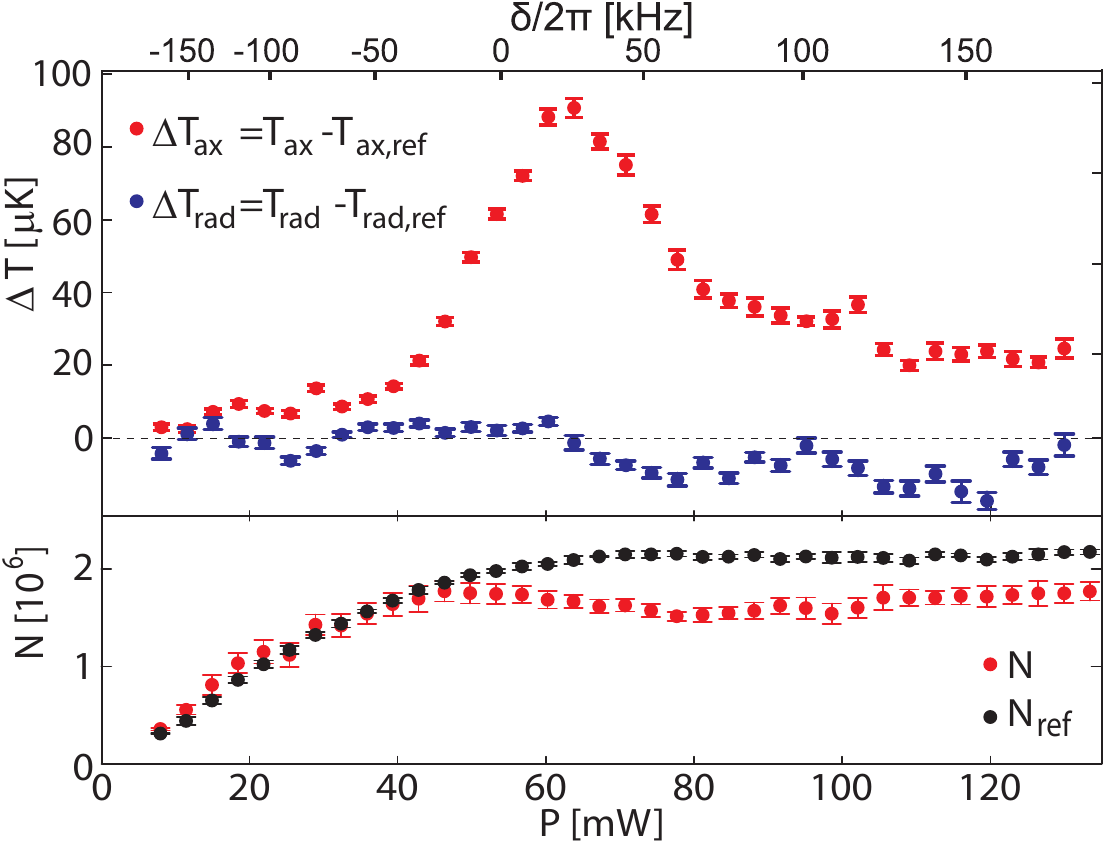}
\caption{\label{fig:paper3} Effect of membrane vibrations on the atoms when laser cooling is off. Top: temperature increase of the atoms along the lattice $\Delta T_{ax}$ and in the radial direction $\Delta T_{rad}$ for a driven membrane with respect to reference measurements for an undriven membrane. 
Bottom: dependence of lattice atom number on $P$, for driven and undriven membrane. 
}
\end{figure}

The resonant coupling should also be visible in the action of the membrane on the atoms. We study this using a different sequence where the membrane is continuously driven at a fixed amplitude of $330$~pm. After loading the lattice, the MOT is switched off and the lattice holds the atoms for an additional $5$~ms. During this time, the membrane motion excites the atomic c.o.m.\ mode in the lattice. Anharmonicity of the lattice couples the c.o.m.\ to other collective modes of the atoms, resulting in heating. 
We determine the axial ($T_{ax}$) and radial ($T_{rad}$) temperatures of the atoms in the lattice from absorption images taken after a time-of-flight of a few ms. 
In Fig.~\ref{fig:paper3} (top), such measurements are shown as a function of $P$. 
We observe a resonant increase in $T_{ax}$ compared to a reference measurement where the membrane is undriven. $T_{rad}$, on the other hand, remains nearly unchanged. 
The shape of the resonance is again influenced by the thermal distribution of the atoms. 
Moreover, we observe that at higher $P$ a fraction of the atoms evaporates from the trap if the membrane is driven (Fig.~\ref{fig:paper3} bottom). This could explain the shift of the resonance with respect to Fig.~\ref{fig:paper4}. 
Quantitative modeling of the data in Fig.~\ref{fig:paper3} is difficult as it would have to account for anharmonic motion, collisions, and evaporation of the atoms. 




In conclusion, we have realized a hybrid optomechanical system composed of ultracold atoms and a membrane. Our observation of backaction of the atoms onto the membrane and the predictions of \cite{optlattmirror} agree remarkably well, suggesting that the theory can be used for extrapolation to optimized parameters. To enhance $g$, large $N$ is favourable. In our current setup, optical access to the MOT chamber is rather limited, and we load only $2 \times 10^{6}$ atoms into the red detuned 1D optical lattice. In a dedicated setup, Raman sideband cooling could be used to prepare up to $3 \times 10^{8}$ atoms in the ground state of a large volume 3D lattice, see \cite{3draman}. In this case, contributions to $\gamma_{at}$ from spatial inhomogeneities and finite $T$ would be much smaller. 
A blue detuned lattice along the membrane direction would suppress trap loss due to light assisted collisions \cite{3draman} while maintaining small laser detuning and power; in the transverse direction the atoms could be confined by a far-detuned 2D lattice \cite{3dlight}. 
In such a setup, the reactive part of the atoms-membrane coupling could be observed as a normal-mode splitting. 
The full quantum theory of our system \cite{optlattmirror} also includes various intrinsic and technical noise sources, such as radiation pressure noise acting on membrane and atoms. 
It shows that the atoms could be used for sympathetic cooling of the membrane to the quantum ground state. 

We acknowledge helpful and inspiring discussions with K. Hammerer, C. Genes, K. Stannigel, M. Wallquist, P. Zoller, M. Ludwig, F. Marquardt, M. Mader, and M. Rakher. Work supported by the Nanosystems Initiative Munich (NIM), the EU project AQUTE, and the NCCR Quantum Science and Technology.


\begin{thebibliography}{27}
\expandafter\ifx\csname natexlab\endcsname\relax\def\natexlab#1{#1}\fi
\expandafter\ifx\csname bibnamefont\endcsname\relax
  \def\bibnamefont#1{#1}\fi
\expandafter\ifx\csname bibfnamefont\endcsname\relax
  \def\bibfnamefont#1{#1}\fi
\expandafter\ifx\csname citenamefont\endcsname\relax
  \def\citenamefont#1{#1}\fi
\expandafter\ifx\csname url\endcsname\relax
  \def\url#1{\texttt{#1}}\fi
\expandafter\ifx\csname urlprefix\endcsname\relax\def\urlprefix{URL }\fi
\providecommand{\bibinfo}[2]{#2}
\providecommand{\eprint}[2][]{\url{#2}}

\bibitem[{\citenamefont{Chu}(1991)}]{Chu1991}
\bibinfo{author}{\bibfnamefont{S.}~\bibnamefont{Chu}},
  \bibinfo{journal}{Science} \textbf{\bibinfo{volume}{253}},
  \bibinfo{pages}{861} (\bibinfo{year}{1991}).

\bibitem{optomech_reviews}
\bibinfo{author}{\bibfnamefont{T.~J.} \bibnamefont{Kippenberg}}
  \bibnamefont{and} \bibinfo{author}{\bibfnamefont{K.}~\bibnamefont{Vahala}},
  \bibinfo{journal}{Science} \textbf{\bibinfo{volume}{321}},
  \bibinfo{pages}{1172} (\bibinfo{year}{2008}), 
\bibinfo{author}{\bibfnamefont{F.}~\bibnamefont{Marquardt}} \bibnamefont{and}
  \bibinfo{author}{\bibfnamefont{S.~M.} \bibnamefont{Girvin}},
  \bibinfo{journal}{Physics} \textbf{\bibinfo{volume}{2}}, \bibinfo{pages}{40}
  (\bibinfo{year}{2009}), 
\bibinfo{author}{\bibfnamefont{I.}~\bibnamefont{Favero}} \bibnamefont{and}
  \bibinfo{author}{\bibfnamefont{K.}~\bibnamefont{Karrai}},
  \bibinfo{journal}{Nat. Photonics} \textbf{\bibinfo{volume}{3}},
  \bibinfo{pages}{201} (\bibinfo{year}{2009}).

\bibitem[{\citenamefont{Chu}(2002)}]{chu_insight}
\bibinfo{author}{\bibfnamefont{S.}~\bibnamefont{Chu}},
  \bibinfo{journal}{Nature} \textbf{\bibinfo{volume}{416}},
  \bibinfo{pages}{206} (\bibinfo{year}{2002}).

\bibitem[{\citenamefont{Bloch}(2005)}]{Bloch2005}
\bibinfo{author}{\bibfnamefont{I.}~\bibnamefont{Bloch}},
  \bibinfo{journal}{Nature Phys.} \textbf{\bibinfo{volume}{1}},
  \bibinfo{pages}{23} (\bibinfo{year}{2005}).
  
\bibitem[{\citenamefont{Murch}(2008)}]{Murch2008}
\bibinfo{author}{\bibfnamefont{K. W.}~\bibnamefont{Murch}} \textit{et al.},
  \bibinfo{journal}{Nature Phys.} \textbf{\bibinfo{volume}{4}},
  \bibinfo{pages}{561} (\bibinfo{year}{2008}).

\bibitem[{\citenamefont{Meiser and Meystre}(2006)}]{prop4}
\bibinfo{author}{\bibfnamefont{D.}~\bibnamefont{Meiser}} \bibnamefont{and}
  \bibinfo{author}{\bibfnamefont{P.}~\bibnamefont{Meystre}},
  \bibinfo{journal}{Phys. Rev. A} \textbf{\bibinfo{volume}{73}},
  \bibinfo{pages}{033417} (\bibinfo{year}{2006}).

\bibitem[{\citenamefont{Genes et~al.}(2008)\citenamefont{Genes, Vitali, and
  Tombesi}}]{prop6}
\bibinfo{author}{\bibfnamefont{C.}~\bibnamefont{Genes}},
  \bibinfo{author}{\bibfnamefont{D.}~\bibnamefont{Vitali}}, \bibnamefont{and}
  \bibinfo{author}{\bibfnamefont{P.}~\bibnamefont{Tombesi}},
  \bibinfo{journal}{Phys. Rev. A} \textbf{\bibinfo{volume}{77}},
  \bibinfo{pages}{050307} (\bibinfo{year}{2008}).

\bibitem{prop7} 
H.~Ian \textit{et al.}, {Phys. Rev. A} \textbf{\bibinfo{volume}{78}},
  \bibinfo{pages}{013824} (\bibinfo{year}{2008}).

\bibitem[{\citenamefont{Bhattacherjee}(2009)}]{prop9}
\bibinfo{author}{\bibfnamefont{A.~B.} \bibnamefont{Bhattacherjee}},
  \bibinfo{journal}{Phys. Rev. A} \textbf{\bibinfo{volume}{80}},
  \bibinfo{pages}{043607} (\bibinfo{year}{2009}).

\bibitem[{\citenamefont{Hammerer
  et~al.}(2009{\natexlab{a}})\citenamefont{Hammerer, Aspelmeyer, Polzik, and
  Zoller}}]{prop10}
\bibinfo{author}{\bibfnamefont{K.}~\bibnamefont{Hammerer}},
  \bibinfo{author}{\bibfnamefont{M.}~\bibnamefont{Aspelmeyer}},
  \bibinfo{author}{\bibfnamefont{E.~S.} \bibnamefont{Polzik}},
  \bibnamefont{and} \bibinfo{author}{\bibfnamefont{P.}~\bibnamefont{Zoller}},
  \bibinfo{journal}{Phys. Rev. Lett.} \textbf{\bibinfo{volume}{102}},
  \bibinfo{pages}{020501} (\bibinfo{year}{2009}{\natexlab{a}}).

\bibitem{Chang09}
Y.~Chang \textit{et al.}, preprint arXiv:0905.0970 (\bibinfo{year}{2009}).

\bibitem{strong_single_atom}
K.~Hammerer \textit{et al.},
{Phys. Rev. Lett.} \textbf{\bibinfo{volume}{103}},
  \bibinfo{pages}{063005} (\bibinfo{year}{2009}{\natexlab{b}}).

\bibitem{Wallquist2010}
M.~Wallquist \textit{et al.},
{Phys. Rev. A} \textbf{\bibinfo{volume}{81}},
  \bibinfo{pages}{023816} (\bibinfo{year}{2010}).

\bibitem{optlattmirror}
K.~Hammerer \textit{et al.},
{Phys. Rev. A} \textbf{\bibinfo{volume}{82}},
  \bibinfo{pages}{021803} (\bibinfo{year}{2010}).

\bibitem[{\citenamefont{Zhang et~al.}(2010)\citenamefont{Zhang, Chen,
  Bhattacharya, and Meystre}}]{prop11}
\bibinfo{author}{\bibfnamefont{K.}~\bibnamefont{Zhang}},
  \bibinfo{author}{\bibfnamefont{W.}~\bibnamefont{Chen}},
  \bibinfo{author}{\bibfnamefont{M.}~\bibnamefont{Bhattacharya}},
  \bibnamefont{and} \bibinfo{author}{\bibfnamefont{P.}~\bibnamefont{Meystre}},
  \bibinfo{journal}{Phys. Rev. A} \textbf{\bibinfo{volume}{81}},
  \bibinfo{pages}{013802} (\bibinfo{year}{2010}).

\bibitem{Hunger2011}
D.~Hunger \textit{et al.}, 
C. R. Physique  (\bibinfo{year}{2011}), doi:10.1016/j.crhy.2011.04.015.

\bibitem{kitching}
Y.-J.~Wang \textit{et al.}, 
{Phys. Rev. Lett.} \textbf{\bibinfo{volume}{97}},
  \bibinfo{pages}{227602} (\bibinfo{year}{2006}).

\bibitem{hunger_prl}
D.~Hunger \textit{et al.}, 
{Phys. Rev. Lett.} \textbf{\bibinfo{volume}{104}},
  \bibinfo{pages}{143002} (\bibinfo{year}{2010}).
  
\bibitem{raithelgoerlitz}
  \bibinfo{author}{\bibfnamefont{G.}~\bibnamefont{Raithel}},
  \bibinfo{author}{\bibfnamefont{W.~D.} \bibnamefont{Phillips}},
  \bibnamefont{and} \bibinfo{author}{\bibfnamefont{S.~L.}
  \bibnamefont{Rolston}}, \bibinfo{journal}{Phys. Rev. Lett.}
  \textbf{\bibinfo{volume}{81}}, \bibinfo{pages}{3615} (\bibinfo{year}{1998});
\bibinfo{author}{\bibfnamefont{A.}~\bibnamefont{G{\"o}rlitz}},
  \bibinfo{author}{\bibfnamefont{M.}~\bibnamefont{Weidem{\"u}ller}},
  \bibinfo{author}{\bibfnamefont{T.~W.} \bibnamefont{H{\"a}nsch}},
  \bibnamefont{and}
  \bibinfo{author}{\bibfnamefont{A.}~\bibnamefont{Hemmerich}},
  \bibinfo{journal}{\textit{ibid.}} \textbf{\bibinfo{volume}{78}},
  \bibinfo{pages}{2096} (\bibinfo{year}{1997}).

\bibitem[{\citenamefont{Grimm et~al.}(2000)\citenamefont{Grimm,
  Weidem{\"u}ller, and Ovchinnikov}}]{Grimm2000}
\bibinfo{author}{\bibfnamefont{R.}~\bibnamefont{Grimm}},
  \bibinfo{author}{\bibfnamefont{M.}~\bibnamefont{Weidem{\"u}ller}},
  \bibnamefont{and}
  \bibinfo{author}{\bibfnamefont{Y.}~\bibnamefont{Ovchinnikov}},
  \bibinfo{journal}{Adv. At. Mol. Opt. Phys.} \textbf{\bibinfo{volume}{42}},
  \bibinfo{pages}{95} (\bibinfo{year}{2000}).

\bibitem[{\citenamefont{Reichel et~al.}(1999)\citenamefont{Reichel, H{\"a}nsel,
  and H{\"a}nsch}}]{mirrormot}
\bibinfo{author}{\bibfnamefont{J.}~\bibnamefont{Reichel}},
  \bibinfo{author}{\bibfnamefont{W.}~\bibnamefont{H{\"a}nsel}},
  \bibnamefont{and} \bibinfo{author}{\bibfnamefont{T.~W.}
  \bibnamefont{H{\"a}nsch}}, \bibinfo{journal}{Phys. Rev. Lett.}
  \textbf{\bibinfo{volume}{83}}, \bibinfo{pages}{3398} (\bibinfo{year}{1999}).

\bibitem{Zwickl08}
B.~M.~Zwickl \textit{et al.}, 
\bibinfo{journal}{Appl. Phys. Lett.}
  \textbf{\bibinfo{volume}{92}}, \bibinfo{pages}{103125}
  (\bibinfo{year}{2008}).

\bibitem{tobepub}
A. J{\"o}ckel \textit{et al.}, \bibinfo{journal}{to be
  published}  (\bibinfo{year}{2011}).

\bibitem[{\citenamefont{Kerman et~al.}(2000)\citenamefont{Kerman, Vuletic,
  Chin, and Chu}}]{3draman}
\bibinfo{author}{\bibfnamefont{A.~J.} \bibnamefont{Kerman}},
  \bibinfo{author}{\bibfnamefont{V.}~\bibnamefont{Vuletic}},
  \bibinfo{author}{\bibfnamefont{C.}~\bibnamefont{Chin}}, \bibnamefont{and}
  \bibinfo{author}{\bibfnamefont{S.}~\bibnamefont{Chu}},
  \bibinfo{journal}{Phys. Rev. Lett.} \textbf{\bibinfo{volume}{84}},
  \bibinfo{pages}{439} (\bibinfo{year}{2000}).

\bibitem{3dlight}
M.~T.~DePue \textit{et al.},
{Phys. Rev. Lett.} \textbf{\bibinfo{volume}{82}},
  \bibinfo{pages}{2262} (\bibinfo{year}{1999}).

\end{thebibliography}
\end{document}